\documentclass[
]{ceurart}

\usepackage{listings}
\usepackage{booktabs,multirow,tabularx}
\usepackage{tikz}
\usepackage{graphicx}
\usetikzlibrary{shapes.geometric, shapes.multipart, arrows.meta, positioning, fit, backgrounds, calc, patterns, shadows}

\usepackage{todonotes}

\begin{document}

\copyrightyear{2026}
\copyrightclause{Copyright for this paper by its authors. Use permitted under Creative Commons License Attribution 4.0 International (CC BY 4.0).}

\conference{UYMS'26: 17. Ulusal Yazılım Mühendisliği Sempozyumu (UYMS), Mayıs 14--16, 2026, Mugla, Turkiye}

\title{An Empirical Evaluation of Using Large Language Models for Automated Model-Based Test Generation
}


\author[1]{Hafize Sanli}[%
orcid=,
email=hafizesanli@posta.mu.edu.tr,
url=,
]
\cormark[1]
\fnmark[1]
\address[1]{Mugla Sitki Kocman University, Mugla, Turkiye}
\address[2]{University of Antwerp, Antwerp, Belgium}

\author[2, 3]{Onur Kilincceker}[%
orcid=,
email=Onur.Kilincceker@uantwerpen.be,
url=,
]
\fnmark[1]

\author[1]{Cihat Cetinkaya}[%
orcid=,
email=cihat.cetinkaya@mu.edu.tr,
url=,
]
\fnmark[1]
\address[3]{Flanders Make vzw, Lommel, Belgium}

\cortext[1]{Corresponding author.}
\fntext[1]{These authors contributed equally.}

\begin{abstract}
Large language models (LLMs) have shown strong potential for automating, analysing, and interpreting software engineering tasks, such as software testing. Model-based testing (MBT) is a software testing technique that poses a broad scalability challenge for industrial adoption. This paper proposes an empirical evaluation of Large Language Models (LLMs) for automated model-based test generation to address scalability challenges, compared with a state-of-the-art model-based testing tool and its built-in algorithms. Our evaluation indicates strong potential to optimize and shorten test paths and step sizes using recent state-of-the-art LLMs on four realistic systems: two web applications and two hardware applications.
\end{abstract}

\begin{keywords}
Empirical Evaluation \sep
  Large Language Models \sep
  Model-Based Test Generation \sep
  Model-Based Testing
\end{keywords}

\maketitle


\noindent{\fontsize{18}{22}\selectfont\sffamily\bfseries Otomatik Model Tabanlı Test Üretimi İçin Büyük Dil Mo\-del\-le\-ri\-nin Kullanımına Dair Ampirik Değerlendirme\par}

\vspace{1em}

\begingroup
    \leftskip = 0.1\textwidth 
    \parindent = 0pt
    
    \noindent \textbf{\sffamily Özet} \par
    \footnotesize 
    \vspace{0.2em}
    \everypar{\parindent=1.5em} 
    \noindent Büyük dil modelleri (LLM), yazılım testi gibi yazılım mühendisliği görevlerini otomatikleştirme, analiz etme ve yorumlama konusunda güçlü bir potansiyel göstermiştir. Model tabanlı test (MBT), endüstriyel benimseme için geniş ölçeklenebilirlik sorunu ortaya koyan bir yazılım test tekniğidir. Bu makale, ölçeklenebilirlik sorunlarını ele almak amacıyla, otomatik model tabanlı test üretimi için Büyük Dil Modellerinin (LLM) deneysel bir değerlendirmesini önermekte ve bunu en son teknolojiye sahip bir model tabanlı test aracı ve yerleşik algoritmalarıyla karşılaştırmaktadır. Değerlendirmemiz, iki web uygulaması ve iki donanım uygulaması olmak üzere dört gerçekçi sistem üzerinde, son teknoloji LLM’ler kullanarak test yollarını ve adım boyutlarını optimize etme ve kısaltma konusunda güçlü bir potansiyel göstermektedir.
    \par

    \vspace{1em}

    \parindent = 0pt
    \noindent \textbf{\sffamily Anahtar Kelimeler} \par
    \vspace{0.2em}
    \noindent Ampirik Değerlendirme, Büyük Dil Modelleri, Model Tabanlı Test Üretimi, Model Tabanlı Test
    \par
\endgroup

\section{Introduction}

Large language models have shown prominent potential in automating, analysing, and interpreting software engineering tasks~\cite{hou2025llms, fan2023large}. They are also actively involved in the software testing process, as part of software engineering activities~\cite{wang2024software, molina2025llmtesting, li2025evaluating, schafer2023empirical}.

Model-based testing (MBT) is a software testing approach in which models are used to describe the expected behavior of the system under test (SUT)~\cite{dalal1999model, utting2010practical}. Automated test generation is a fundamental part of model-based testing where testers can generate abstract test cases (offline mode) or directly run the automated test cases on the system (online mode) under test. It has evolved significantly, with extensive research expanding its application across web, mobile, and embedded systems~\cite{garousi2021model, karlsson2021model, zafar2021model}. Numerous studies and taxonomies have categorized a broad range of MBT tools and techniques~\cite{utting2016recent, li2017survey, gurbuz2018model, ahmad2019model}.

Scalability poses a broad challenge for the industrial adoption of MBT~\cite{dalal1999model, hemmati2010enhanced, hemmati2010reducing, koroglu2025towards}. Common practice suggests that longer tests are preferable~\cite{arcuri2010longer}; however, prior work on test reduction has demonstrated that more compact test cases achieving the same coverage goals enhance scalability~\cite{rothermel2002empirical, coutinho2016analysis}.

In this study, we empirically assess Large Language Models (LLMs) for automated model-based test generation to address scalability challenges. Specifically, we investigate how well LLMs perform in producing shorter tests compared to the baseline state-of-the-art model-based testing tool (GraphWalker\footnote{See~\url{https://graphwalker.github.io}.}) with its built-in algorithms.

Based on all the previous discussion, we present the following contributions to the literature.
\begin{itemize}
    \item We present an automation pipeline (called LLM4MBT) for automating the process of model-based test generation via LLMs.
    \item We introduce an empirical evaluation on the usage of recent LLMs for automated model-based test generation.
    \item We perform experiments on four realistic SUTs: two web applications (Parabank and Testinium) and two hardware applications (TLC and RISC-V) for answering 5 research questions and show that LLM4MBT feasible.
    \item We publicly share our dataset, LLM4MBT pipeline implementation and replication instruction at \url{https://github.com/hafizesanli/LLM4MBT} (Version 1).
\end{itemize}

The rest of this paper is organized  as starting with the approach we propose in Section \ref{approach} and we present the research questions and evaluation setup in Section \ref{rqs} and our empirical evaluation in Section \ref{eval}. We describe the potential threats to validity including their mitigation methods in Section \ref{threats}. We discuss the related work in Section \ref{related} and conclude the paper and share the future work in Section \ref{conclusion}.

\section{Approach}
\label{approach}

Our work presents a systematic framework for evaluating the capability of Large Language Models (LLMs) to generate valid model-based test suites for GraphWalker models. Unlike traditional automated test generation approaches that rely on algorithmic traversal or heuristic-based exploration, we investigate whether LLMs can leverage their learned knowledge of software testing principles and GraphWalker specifications to produce test paths that achieve specific coverage criteria.

Our methodology evaluates five state-of-the-art LLMs (GPT-5.1, GPT-5.2, Claude Opus 4.5, Claude Sonnet 4.5, and Gemini 2.5 Pro) against four GraphWalker models of escalating complexity. These models range from a simple Traffic Light Controller (TLC) to a highly complex enterprise Test Management Platform (Testinium) containing 129 vertices and 259 edges. The primary objective is to determine if a single, well-structured prompt can elicit syntactically and logically valid test suites that meet predefined coverage goals.

\begin{figure}[htbp]
  \centering
  \includegraphics[width=0.8\textwidth]{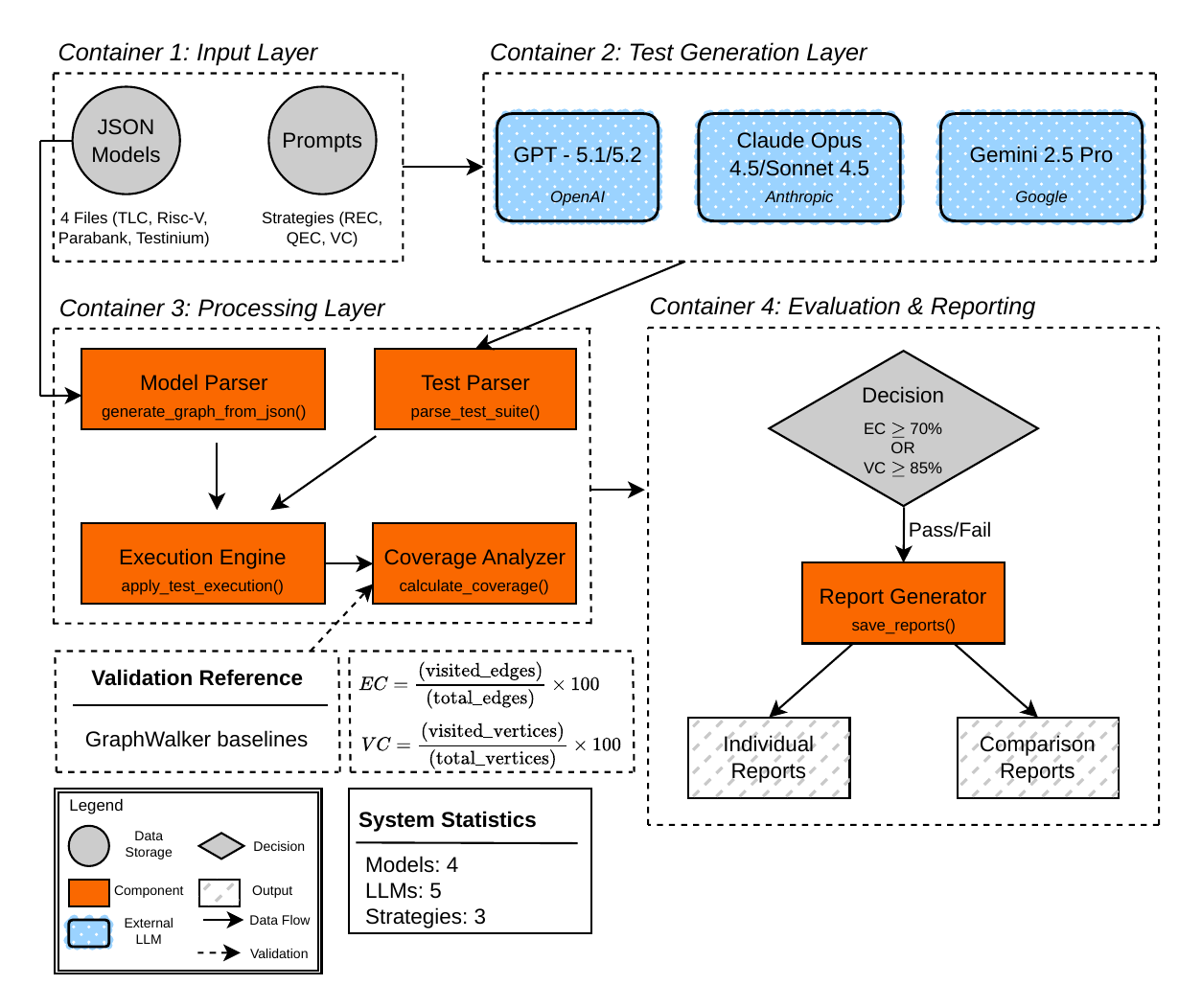}
  \caption{LLM4MBT pipeline architecture}
  \label{fig:llm4mbt}
\end{figure}

\subsection{LLM4MBT Pipeline Architecture}

Our proposed LLM4MBT framework implements a multi-layered pipeline architecture for systematic evaluation of LLMs in the context of model-based test generation. Figure~\ref{fig:llm4mbt} illustrates the complete system architecture, which consists of four distinct containers following the C4 model architecture principles~\cite{vazquez2020c4}: Input Layer, Test Generation Layer, Processing Layer, and Evaluation \& Reporting Layer.

\textbf{Input Layer:}
The pipeline initiates with the Input Layer (Container 1 in Figure~\ref{fig:llm4mbt}), which manages all input data sources for the experimental framework. This layer comprises two primary components: (i) \textit{Graph Models}, consisting of four GraphWalker models represented in JSON format---TLC (10 vertices, 18 edges), Risc-V (16 vertices, 36 edges), Parabank (75 vertices, 144 edges), and Testinium (129 vertices, 259 edges)---which vary significantly in complexity to enable comprehensive evaluation across different scales; and (ii) \textit{Prompts}, stored in \texttt{prompts.json}, which contains manually crafted prompt templates for three distinct path generation strategies: Random Edge Coverage (REC), Quick Random Edge Coverage (QEC), and Vertex Coverage (VC). These input components are distributed to subsequent layers to facilitate systematic test generation across all experimental conditions.

\textbf{Test Generation Layer:}
The Test Generation Layer (Container 2 in Figure~\ref{fig:llm4mbt}) represents the core experimental subjects of our evaluation framework. This layer incorporates five state-of-the-art Large Language Models as external systems: GPT-5.1 and GPT-5.2 from OpenAI, Claude Opus 4.5 and Claude Sonnet 4.5 from Anthropic, and Gemini 2.5 Pro from Google. Each LLM receives identical inputs consisting of the graph models and strategy-specific prompts, ensuring experimental consistency. The combinatorial design of this layer, 5 LLMs $\times$ 3 strategies $\times$ 4 models, generates 60 distinct test suites, which are systematically organized into strategy-specific directories (\texttt{REC-*/}, \texttt{QEC-*/}, \texttt{VC-*/}) for subsequent processing and analysis.

\textbf{Processing Layer:}
The Processing Layer (Container 3 in Figure~\ref{fig:llm4mbt}) implements the core computational logic for test execution and coverage analysis. This layer consists of four interconnected components:

Model Parser: Implemented in \texttt{graph\_conversions.py} through the \texttt{generate\_graph\_from\_graphwalker\_json()} function, this component transforms JSON model specifications into internal graph representations, extracting vertices, edges, and their relationships for subsequent processing.
    
Test Parser: Located in \texttt{main.py} and implemented via \texttt{parse\_test\_suite\_from\_file()}, this component processes LLM-generated test suites, extracting individual test cases in GraphWalker's raw format where each element is represented as a JSON object containing the \texttt{currentElementName} field.
    
Execution Engine: Responsible for path validation and test execution, this component (\texttt{apply\_test\_execution\_on\_model()} in \texttt{main.py}) validates whether generated paths conform to the model's graph structure. It incorporates intelligent fallback mechanisms (\texttt{find\_fallback\_vertex()}) to handle invalid transitions, thereby enabling fair evaluation even when LLMs generate partially incorrect test sequences.
    
Coverage Analyzer: Implemented through \texttt{calculate\_coverage()} in \texttt{graph\_conversions.py}, this component computes two fundamental metrics: Edge Coverage (EC) and Vertex Coverage (VC), defined as:
\begin{equation}
\label{equ1}
EC = \frac{|E_{visited}|}{|E_{total}|} \times 100, VC = \frac{|V_{visited}|}{|V_{total}|} \times 100
\end{equation}
where $E_{visited}$ and $V_{visited}$ represent the sets of traversed edges and vertices, respectively.

To ensure validity, the Coverage Analyzer's computations are validated against reference baselines generated by GraphWalker's native test generation engine, as indicated by the dotted validation arrow in Figure~\ref{fig:llm4mbt}.

\textbf{Evaluation \& Reporting Layer:}
The final layer (Container 4 in Figure~\ref{fig:llm4mbt}) implements the evaluation logic and report generation mechanisms. The \textbf{Decision} component applies predefined success criteria to classify each test suite as PASS or FAIL based on the condition: $EC \geq 70\%$ OR $VC \geq 85\%$. These thresholds were established based on industry standards for model-based testing adequacy. 

The Report Generator component (\texttt{save\_single\_test\_report()} and \texttt{save\_comparison\_reports()} in \texttt{main.py}) produces two categories of outputs: (i) \textit{Individual Reports}, stored in \texttt{test\_reports/}, which document the performance of each test suite independently in both JSON and readable text formats; and (ii) \textit{Comparison Reports}, stored in \texttt{comparison\_reports/}, which aggregate results across all LLMs and strategies to facilitate comparative analysis. These reports provide comprehensive documentation of coverage metrics, execution statistics, and pass/fail outcomes for all 60 experimental conditions.

The modular architecture illustrated in Figure~\ref{fig:llm4mbt} enables systematic evaluation while maintaining clear separation of concerns: data management (Input Layer), test generation (Test Generation Layer), execution and measurement (Processing Layer), and result aggregation (Evaluation \& Reporting Layer). This design facilitates reproducibility and extensibility for future research incorporating additional models, LLMs, or coverage strategies.

\subsection{Algorithm Details}
The evaluation framework employs a rigorous validation algorithm to assess the quality of the LLM generated paths. The core logic centers on Path Validation and Coverage Analysis.

\textbf{Execution Logic and Fallback Mechanism:}
The execution engine iterates through each transition in the generated test suite. For every step, the algorithm verifies the existence of an edge \(E\) between the current vertex \(V_n\) and the target vertex \(V_{n+1}\). To account for the stochastic nature of LLMs, we implement a fallback recovery mechanism:\par
If an invalid transition is encountered, the engine does not terminate the entire suite. Instead, it attempts to restart the traversal from a known stable vertex (e.g., "start" or "v\_Login"). This allows for the measurement of "partial success" and ensures that valid segments of a suite are still credited toward total coverage.

\textbf{Coverage Metrics:}
Our framework calculates two primary metrics to assess the quality of LLM-generated test suites, with success criteria defined as Edge Coverage $\geq 70\%$ OR Vertex Coverage $\geq 85\%$:

Edge Coverage (EC): Quantifies the proportion of unique edges traversed during test execution relative to the total number of edges in the model. It was formally given in (\ref{equ1}), where $E_{visited}$ denotes the set of edges traversed by the test suite and $E_{total}$ represents all edges defined in the model.

Vertex Coverage (VC): Measures the proportion of unique vertices visited during test execution relative to the total number of vertices in the model. This metric is defined in (\ref{equ2}), where $V_{visited}$ represents the set of vertices visited by the test suite and $V_{total}$ denotes all vertices in the model.

These metrics provide complementary perspectives on test adequacy: edge coverage emphasizes transition testing (interactions between states), while vertex coverage focuses on state space exploration. The dual criteria ensure that test suites achieving either high transition coverage or comprehensive state coverage are considered successful, accommodating different testing priorities and model characteristics. The Coverage Analyzer component validates these computations against reference baselines generated by GraphWalker's native test generation engine to ensure measurement accuracy.

\section{Research Questions and Evaluation Setup}
\label{rqs}

This section presents our research questions, motivation, and experimental setup for conducting the experiments to address them.

\subsection{Research Questions}

Our empirical evaluation is designed to address the following research questions (RQs).

\textbf{Evaluation of Coverage} 

\noindent\fbox{%
    \parbox{\textwidth}{%
        \textbf{RQ1:} To what extent does LLM4MBT achieve vertex and edge coverage in model-based test generation?
    }%
}
In GraphWalker, we set coverage ratios to 100\% for edges and vertices. In the prompt for utilized LLMs, we also set a 100\% coverage target; however, due to failing test paths, we typically do not expect to achieve 100\% coverage for any LLM. 

\textbf{Comparative Analysis}

\noindent\fbox{%
    \parbox{\textwidth}{%
        \textbf{RQ2:} How does the coverage of LLM4MBT-generated test steps compare to the GraphWalker baseline?
    }%
}
As noted in RQ1, although we do not expect 100\% coverage, the overall test step length is expected to be shorter than that of the GraphWalker baseline. Our motivation is to highlight those comparisons in this RQ. 




\textbf{Ablation Study} 

\noindent\fbox{%
    \parbox{\textwidth}{%
        \textbf{RQ3:} How do specific prompt components influence the effectiveness of LLM4MBT-generated test steps?
    }%
}

As part of our ablation study, we aim to examine how LLMs behave under different prompts and to quantify their individual contributions to overall performance.

\textbf{Model Dependency} 

\noindent\fbox{%
    \parbox{\textwidth}{%
        \textbf{RQ4:} To what degree does the coverage achieved by LLM4MBT correlate with the capabilities of the underlying LLM?
    }%
}

The experimental results may vary with the five state-of-the-art LLMs used, and we aim to quantify the extent to which these results depend on the underlying LLMs.

\subsection{Evaluation Setup}

Our evaluation encompasses 4 realistic GraphWalker models across 2 domains: embedded systems and web applications. A traffic light controller (TLC) \footnote{See \url{https://github.com/kilincceker/MBIT4HW}.} is a simple embedded system and is a finite state machine representing a traffic light controller with states q0-q8, and a start vertex~\cite{kilinccceker2018regular, kilincceker2022model}. RISC-V\footnote{See \url{https://github.com/openhwgroup/cv32e40p}.} is an embedded CPU model with various instruction sets and pipelines~\cite{kilinccceker2018regular, kilincceker2022model}. Parabank\footnote{See \url{https://github.com/parasoft/parabank}.} is a multi-module banking web application with Login, Register, and AccountOverview modules~\cite{koroglu2025towards}. Testinium\footnote{See \url{https://github.com/vgarousi/MBTofTestinium}.} is an enterprise test management web platform with multiple modules: Login, Dashboard, Reports, Projects, Scenarios, and Suites~\cite{garousi2021model,garousi2024coverage}.

As provided in Table \ref{tab:llm_characteristics}, our evaluation comprises five state-of-the-art Large Language Models as external systems: GPT-5.1 and GPT-5.2 from OpenAI, Claude Opus 4.5 and Claude Sonnet 4.5 from Anthropic, and Gemini 2.5 Pro from Google.

\begin{table*}[t]

\centering
\scriptsize
\setlength{\tabcolsep}{5.0pt}
\renewcommand{\arraystretch}{1.2}

\caption{Characteristics of Large Language Models used in the experiment}
\label{tab:llm_characteristics}

\begin{tabularx}{\textwidth}{l r r l X}
\toprule
\multirow{1}{*}{Model} &
\multirow{1}{*}{Year} & 
\multirow{1}{*}{Parameters} & 
\multirow{1}{*}{Provider} & 
\multirow{1}{*}{Features} \\

\midrule
GPT-5.1 & 2025 & Not disclosed & OpenAI & 

Multimodal (text+image);
adaptive reasoning; 
improved conversational tone \\

\midrule
GPT-5.2 & 2025 & Not disclosed & OpenAI & 

Multimodal (text+image);
enhanced reasoning (Thinking 
mode); improved coding and 
spreadsheet capabilities\\

\midrule
Claude Opus 4.5 & 2025 & Not disclosed & Anthropic & 

Multimodal (text+image+code); 
frontier reasoning; advanced 
coding; effort parameter;
token-efficient \\

\midrule
Claude Sonnet 4.5 & 2025 & Not disclosed & Anthropic & 

Multimodal (text+image+code); 
best coding performance; 
computer use; agentic tasks;
improved reasoning \\

\midrule
Gemini 2.5 Pro & 2025 & Not disclosed & Google & 

Multimodal (text+image+audio
+video); enhanced reasoning;
Deep Think mode; 1M token
context window \\

\bottomrule

\end{tabularx}
\end{table*}

\section{Evaluation Results}
\label{eval}

In this section, we answer the four RQs through experiments on four realistic GraphWalker models across five state-of-the-art LLMs given in Table \ref{tab:llm_characteristics}.

\subsection{RQ1: LLM4MBT’s Coverage}

To evaluate the coverage effectiveness of LLM4MBT, we conducted experiments across four different model-based testing projects (Traffic Light Controller(TLC), RISC-V, Parabank, and Testinium), comparing it against GraphWalker's baseline strategies.\par
As shown in Table \ref{tab:llm4mbt-graphwalker}, LLM4MBT achieved 100.0\% average vertex coverage and 96.3\% average edge coverage, demonstrating comprehensive coverage metrics.

Table \ref{tab:llm_leaderboard} shows the varying performance among LLM providers, and Table \ref{tab:llm_leaderboard} shows the average performance of all LLMs. These results indicate that LLM4MBT can generate comprehensive test suites with significantly greater efficiency than traditional walk-based approaches, while achieving coverage levels comparable to or exceeding those of baseline tools. Table \ref{tab:llm_leaderboard} further demonstrates that this efficiency pattern holds across different LLM providers, with all models requiring significantly fewer steps than GraphWalker while achieving comparable coverage.


\begin{table*}[t]

\centering
\scriptsize
\setlength{\tabcolsep}{5.0pt}
\renewcommand{\arraystretch}{1.05}

\caption{Coverage and test outcomes for \textsc{LLM4MBT} and GraphWalker}
\label{tab:llm4mbt-graphwalker}

\begin{tabularx}{\textwidth}{l r r r r r r r r r}
\toprule
\multirow{2}{*}{Project} &
\multicolumn{6}{c}{\textsc{LLM4MBT (Claude Opus 4.5-}Quick Random)} &
\multicolumn{3}{c}{GraphWalker(Test Steps-Edge)} \\
\cmidrule(lr){2-7}\cmidrule(lr){8-10}
& Total & Passing (\%) & Vertex & Edge & Test Steps (\#) & Contribution (\%) & Random & Quick Random\\
\midrule
Traffic Light Controller & 46   & 44 (95.6\%)   & 100.0\%        & 94.4\%   & 89  & 18 (16.8\%)    & 2809 & 107 \\
\midrule
RISC-V Controller       & 54    & 54 (100.0\%)   & 100.0\%          & 100.0\%    & 105      & 9 (7.8\%)    & 532 & 114 \\
\midrule
Parabank             & 225 & 46 (20.4\%) & 100.0\% & 100.0\% & 351 & 376 (51.7\%) & 3523 & 727 \\
\midrule
Testinium             & 358 & 358 (100.0\%) & 100.0\% & 91.0\% & 657 & 492 (42.8\%) & 20799 & 1149 \\
\midrule
\textbf{Average} &  & 79.0\% & \textbf{100.0\%} & \textbf{96.3\%} & \textbf{300} & 29.7\% & \textbf{6916} & \textbf{524}\\
\bottomrule

\end{tabularx}

\end{table*}

\subsection{RQ2: LLM4MBT vs. GraphWalker}

While RQ1 established LLM4MBT's coverage capabilities, RQ2 examines the efficiency trade-off between coverage and test suite size relative to GraphWalker. Table \ref{tab:llm4mbt-graphwalker} reveals a significant efficiency advantage for LLM4MBT. Notably, LLM4MBT resulted in only about 300 average test steps, compared with GraphWalker's 524 steps (Quick Random), representing a substantial efficiency improvement over random approaches while maintaining comparable coverage. Moreover, Table \ref{tab:llm4mbt-graphwalker} shows that our best performing LLM and its algorithm (Claude Opus 4.5-Quick Random) contribute with an average of 29.7\% to GraphWalker's best performing algorithm (quick random) based on the given formula in (\ref{equ4}). Also, once the model size grows, the contribution gaps and percentage increases. 

\begin{equation}
    \label{equ4}
    Contribution = \frac{|Test Steps_{GraphWalker}| - |Test Steps_{LLM4MBT}|}{|Test Steps_{GraphWalker}|} \times 100
\end{equation}

Achieving 96.3\% average edge coverage with only 300 average test steps, compared to GraphWalker's 100.0\% coverage requiring 6916 steps (Random) or 524 steps (Quick Random). This represents a 23.0x reduction in test suite size relative to the Random and a 1.7x reduction relative to Quick Random, while maintaining near-complete coverage.

\subsection{RQ3: Prompt Effect}
As part of our ablation study, we examined how different prompt structures and components affect LLM behavior and their contribution to test generation performance. Analysis of 'prompts.json' reveals critical prompt engineering insights such as Role Definition, Coverage Strategy Specification, Output Format Example, and Constraint Instructions. \par
\textbf{Terminology Precision:} Using "quick\_random" instead of "quick random" significantly improved success rates across all LLMs. Initial attempts with "quick random" frequently failed or produced incorrect outputs.\par
\textbf{Provider Specific Behavior:} GPT models initially attempt to generate Python scripts rather than direct test suites, requiring explicit negative instructions. Claude models handled the task more directly without such constraints.\par
\textbf{Instruction Clarity:} Adding explicit constraints prevented file copying behavior and improved generation quality, particularly for GPT 5.1 which showed higher sensitivity to instruction ambiguity.\par
\textbf{Model Batching:} Generating multiple models in a single prompt worked reliably for Claude models but caused failures with GPT models, necessitating per model prompts for consistent results.
The 'prompts.json' in our dataset (publicly available at \url{https://github.com/hafizesanli/LLM4MBT}) contains those prompt engineering insights.

\subsection{RQ4: Effect of Different LLMs}
We evaluated five state-of-the-art LLMs to quantify how model capabilities affect test generation performance. Table \ref{tab:llm_leaderboard} shows significant performance variations across LLM providers. For Quick Random Edge Coverage, Claude Opus 4.5 achieved the most consistent high performance with 100\% edge coverage scores on RISC-V and Parabank; GPT 5.1 demonstrated 100\% edge coverage with RISC-V performance but showed variability on complex models. Gemini 2.5 Pro, demonstrating the highest variance, achieved 100\% edge coverage in the TLC and Parabank models, but completely failed on RISC-V. For Random Edge Coverage,  Claude models and GPT 5.2 achieved 100.0\% coverage on TLC and RISC-V. Claude Sonnet 4.5 showed superior performance on complex models, whereas Gemini 2.5 Pro again showed high variability.\par

\begin{table*}[t]

\centering
\scriptsize
\setlength{\tabcolsep}{5.0pt}
\renewcommand{\arraystretch}{1.05}

\caption{Model Capability Leaderboard: Comparative Analysis of LLM4MBT Coverage}
\label{tab:llm_leaderboard}

\begin{tabularx}{\textwidth}{l r r r r r r r r r}
\toprule
\multirow{2}{*}{Model Engine} &
\multirow{2}{*}{Provider} &
\multicolumn{4}{c}{Quick Random Edge Coverage} & \multicolumn{4}{c}{Random Edge Coverage} \\
\cmidrule(lr){3-6}\cmidrule(lr){7-10}
& & TLC & Risc-V & Parabank & Testinium & TLC & Risc-V & Parabank & Testinium\\
\midrule
Claude 4.5 Opus        & Anthropic    & 94.44\%   & 100.00\%          & 100.00\%    & 91.09\% & 100.00\% & 100.00\% & 0.00\% & 90.70\%\\
\midrule
Claude 4.5 Sonnet           & Anthropic & 94.44\% & 100.00\% & 84.03\% & 89.15\% & 100.00\%& 100.00\%& 97.22\%& 89.15\% \\
\midrule
GPT-5.2 & OpenAI   & 61.11\%   & 100.00\%        & 79.86\%   & 60.08\%  & 100.00\%& 100.00\%& 74.31\%& 36.43\% \\
\midrule
GPT-5.1             & OpenAI & 100.00\% & 100.00\% & 79.86\% & 88.37\% & 94.44\%& 100.00\%& 73.61\%& 36.43\% \\
\midrule
Gemini 2.5 Pro            & Google & 100.00\% & 0.00\% & 100.00\% & 86.82\% & 94.44\%& 0.00\%& 31.94\%& 0.00\%  \\
\midrule
\textit{Average} &  & \textit{89.99\%} & \textit{80.00\%} & \textit{88.75\%} & \textit{83.10\%} & \textit{97.77\%} & \textit{80.00\%} & \textit{55.41\%} & \textit{50.54\%} \\
\bottomrule

\end{tabularx}

\end{table*}

Anthropic's Claude models demonstrated the most reliable and consistent performance across all models and strategies. OpenAI's GPT models showed moderate consistency with better performance on simpler models. Google's Gemini 2.5 Pro exhibited model-specific strengths but lacked generalization, suggesting a strong correlation between LLM reasoning capabilities and test generation effectiveness. These results indicate that LLM architectural differences and training paradigms (Table \ref{tab:llm_characteristics}) significantly affect test-generation quality, with reasoning consistency being more critical than raw model size for MBT tasks. 

\section{Threats To Validity}
\label{threats}
This section presents potential threats to our empirical evaluation and outlines strategies to mitigate them.

\textbf{Internal Validity:}
Due to the generative nature of LLMs (Claude, GPT, and Gemini), the same prompt can yield different test paths across runs. We plan to mitigate this by multiple iterations/seeds and use a specific temperature setting, e.g., $T=0$. However, because our results depend on GitHub Copilot's capabilities, we were unable to specify those temperature parameters in our current work.

The results from GraphWalker depend heavily on the traversal strategy used (e.g., Random, Quick Random). Therefore, due to the random effect, each run yields different test paths. LLM outputs are inherently stochastic; running multiple trials for each configuration would yield more reliable results. As future work, we will address these by performing multiple iterations (at least 30) or using seeds.

As demonstrated in RQ4, even minor variations in prompt structure can substantially influence coverage. However, these findings may be particular to the prompt engineering strategies we used. To mitigate this, we conducted an ablation study as part of our empirical evaluation with five state-of-the-art LLMs.

\textbf{Construct Validity:}
Even though high vertex and edge coverage are standard MBT performance metrics, they do not always correlate with the capability to find real-world software bugs. A test suite could achieve 100\% coverage without catching a logic flaw if the oracles are weak. To address this, we plan to use mutation testing at the model and code levels to assess their correlation~\cite{belli2016model, kilincceker2021model, kilincceker2022model}. Therefore, we will integrate the test execution phase into our existing pipeline architecture.

\textbf{External Validity:}
We evaluated four different GraphWalker models (Traffic Light, RISC-V, Parabank, Testinium). These results may not generalize to significantly larger industrial systems with thousands of states or highly non-deterministic behavior. For addressing this, we selected the GraphWalker models from two distinguishable domains (hardware and web applications)

The effectiveness of LLM4MBT may vary across different modeling languages (e.g., Statecharts, Petri Nets, BPMN). To mitigate this threat, we employed a state-of-the-art model-based testing tool, GraphWalker, widely used in academia and industry.

\section{Related Work}
\label{related}

The role of Large Language Models (LLMs) in software engineering has shifted from simple code assistants to strategic partners active throughout all phases of the Software Development Life Cycle (SDLC). Significant productivity gains are being achieved by LLMs in software testing, where the field is shifting from manual scripting toward autonomous test agents and intelligent oracles. A pivotal empirical evaluation on the use of LLMs for automated unit test generation was conducted by \cite{schafer2023empirical}, in which it was demonstrated that coverage targets comparable to traditional tools could be achieved while identifying complex edge cases. The prominent potential of LLMs in automating, analyzing, and interpreting software engineering tasks was highlighted in their work.
Beyond simple script generation, LLMs are now utilized as decision-makers to verify whether a system behaves correctly. This is supported by empirical studies in \cite{li2025evaluating}, which show that LLM-based approaches are highly effective at detecting critical bugs and perform as well as traditional tools in achieving coverage targets. The practical utility of LLMs has been empirically validated in specialized domains; for instance, the successful automation of a complete software testing process in the automotive industry was demonstrated in \cite{wang2025automating}. It has been shown that end-to-end testing life cycles in complex, safety-critical environments can be handled by LLMs.

To the best of our knowledge, no study has systematically evaluated the latest Large Language Models (LLMs) for automated Model-Based Testing (MBT) across both web and hardware applications using graph-based specifications. While the efficacy of LLMs has been demonstrated in unit testing and general industrial processes, their ability to navigate complex graph structures to achieve optimized coverage remains underexplored. In this study, the LLM4MBT pipeline is introduced to bridge this gap. It is demonstrated that near-complete coverage can be maintained with significantly fewer test steps than traditional algorithms, thereby directly addressing the scalability challenges inherent in MBT.

\section{Conclusion}
\label{conclusion}

  Large language models have shown strong potential for software engineering tasks, particularly software testing. Model-based testing (MBT) is a software testing technique. To address the broad scalability challenge for industrial adoption of MBTs, our paper presents an empirical evaluation of Large Language Models (LLMs) for automated model-based test generation, compared with a state-of-the-art model-based testing tool (GraphWalker) and its built-in algorithms (random and quick random for edge and vertex coverage settings). Our evaluation indicates strong potential to optimize and shorten test paths and step sizes using the recent five state-of-the-art LLMs (GPT-5.1, GPT-5.2, Claude Opus 4.5, Claude Sonnet 4.5, and Gemini 2.5 Pro) against four GraphWalker models (two web applications (Parabank and Testinium) and two hardware applications (TLC and RISC-V) ) of escalating complexity. 


\section*{Data availability} 
The dataset, LLM4MBT pipeline implementation, and replication instructions are available at \url{https://github.com/hafizesanli/LLM4MBT} (Version 1).

\section*{Declaration on Generative AI}
During the preparation of this work, the author(s) utilized Grammarly and Bing Microsoft Translator to check grammar and spelling. The author(s) reviewed and edited the content as needed and assumes full responsibility for the content of the publication.

\bibliography{bibfile}

\end{document}